\documentclass[final]{elsart}

\usepackage[latin1]{inputenc}
\usepackage{amsmath} 
\usepackage{amssymb}
\usepackage{graphicx} 
\usepackage{dcolumn} 
\usepackage{bm} 
\usepackage[dvips]{color}

\begin{document}

\begin{frontmatter}

\title{Quantification of Continuous Variable Entanglement with only Two
Types of Simple Measurements}




\author{Gustavo Rigolin},
\ead{rigolin@ifi.unicamp.br}
\author{Marcos C. de Oliveira}
\ead{marcos@ifi.unicamp.br}

\address{
Instituto de F\'{\i}sica Gleb Wataghin,
Universidade Estadual de Campinas, Unicamp,
13083-970, Campinas, S\~ao Paulo, Brasil}


\begin{abstract}
Here we propose an experimental set-up in which it is possible to
obtain the entanglement of a two-mode Gaussian state, be it pure or mixed,
using only simple linear optical measurement devices.
After a proper unitary manipulation of the two-mode Gaussian state
only number and purity measurements of just one of the modes suffice to give
us a complete and exact knowledge of the state's entanglement.
\end{abstract}


\end{frontmatter}



\section{Introduction}

Quantum theory of information offers, in principle, a plethora of resources
for the processing of computational and informational tasks in an efficient
way, otherwise unattainable in the classical
world \cite{Ben00,Bra05,Ade07,eisert1}.
Initially, all the quantum communication protocols were
developed for discrete systems (qubits) and later an equivalent formulation
for the quantum continuous variable (CV) setting was developed. CV-systems
 are described by canonical conjugate
operators like position and momentum or the quadrature amplitudes of
the quantized electromagnetic field. The main reason for the development of
a CV quantum information theory are of a practical order: the crucial
steps in quantum communication protocols are relatively simple to implement
via the available experimental techniques of quantum optics \cite{Bra05}.
Some of these tasks,
e.g. quantum cryptography \cite{Eke91,Coh97,Per00},
superdense coding \cite{Wie92,Ban99,Bra00}
and teleportation \cite{Ben93,Vai94,Bra98},
require entanglement as a key ingredient for their successful
implementation.  Therefore, a great deal of effort has been put into the
qualitative and quantitative study of entanglement. However, it is still
a difficult problem to obtain in a simple experimental fashion the degree
of entanglement of a quantum system. Practically all the useful
entanglement measures
developed so far can only be computed with a complete knowledge of the
quantum state, whose reconstruction is not a trivial
experimental task.

A direct scheme for the measurement of the
entanglement of a pair of pure qubits, without state reconstruction,
was recently proposed \cite{Min05}
and experimentally realized \cite{Wal06}.
For continuous variable systems (CV-systems), being Gaussian states  a famous
example which are efficiently generated in the 
laboratory \cite{Bra05,eisert1}, 
interesting
experimental proposals for the measurement of entanglement were presented
\cite{Ade04a,Ade04b,Fiu04} and implemented \cite{Gia04} recently. In
particular, it was shown that by measuring the marginal (local)
and global purities of a two-mode Gaussian state one is able to
estimate the state negativity \cite{Wer02}, although its
exact value cannot be achieved via this procedure.
The negativity, together with the entanglement of formation \cite{Gie03}, are
the most important entanglement measures for Gaussian states, where the
latter is only analytically computable for symmetric Gaussian states.

Here we present a simple experimental scheme allowing the measurement of 
the exact value of the entanglement of formation for an arbitrary symmetric
two-mode Gaussian state. In fact, the scheme can also be adapted to obtain
the exact value of the negativity for  
an arbitrary two-mode Gaussian state as well, being it
symmetric or not. The only prior knowledge needed for entanglement
quantification, which is also a requisite of all previous schemes, is upon the
Gaussian nature of the two-mode state.
Whether it is a pure or a mixed state is irrelevant for our purposes.
Furthermore, even if we do not know that we are dealing with a Gaussian state
our method can be used to experimentally apply the Simon separability
test \cite{Sim00}, which is a sufficient condition for the existence of
CV-entanglement.

A most remarkable feature of our scheme, which differentiates it from
previous proposals,
is that all the information relevant to the
\textit{exact} quantification
of the entanglement of the state can be obtained, after a non-local unitary
operation on the two-modes, via \textit{local projective measurements}.
After combining the
two-modes in a beam splitter (non-local unitary operation) the entanglement
can be quantified solely by
measurements of the purity and the number of photons of just one of
the modes. Therefore, all the information
concerning the entanglement of the
two-modes is transferred to local properties of one of the modes
\textit{after} a non-local unitary operation.
It is in this sense that the entanglement can be seen as
determined by two
single types of local projective measurements with no need for quantum state
reconstruction or, equivalently, without the knowledge of the two-mode 
covariance matrix.

\section{Entanglement of a two-mode Gaussian state}

A two-mode Gaussian state $\rho$ is completely cha\-rac\-te\-rized by its
covariance matrix, with elements
$\gamma_{ij}$ $=$ $\langle R_i R_j + R_j R_i\rangle_{\rho}$
$-$ $2$$\langle R_i\rangle_{\rho}$$\langle R_j \rangle_{\rho}$,
where $\langle R_i \rangle_{\rho}$
is the quantum expectation value of the observable $R_i$ and
$(R_1,R_2,R_3,R_4)$ $=$ $(X_1,P_1,X_2,P_2)$ are the quadratures of the
electromagnetic field modes \cite{footnote1}.
Any covariance matrix can be brought via local
symplectic transformations, i.e. without affecting its entanglement content,
to the standard form \cite{Sim00}
\begin{equation}
\gamma =
\left(
\begin{array}{cc}
A & C_{\gamma} \\
C_{\gamma}^{T} & B
\end{array}
\right),
\label{simon-standard}
\end{equation}
where $A, B$, and $C_{\gamma}$ are $2\times 2$ real diagonal matrices given as
\cite{Gie03}
$A=\mbox{diag}(n,n)$, $B=\mbox{diag}(m,m)$, and
$C_{\gamma}=\mbox{diag}(k_x,-k_p)$. It can be shown \cite{Sim00} that a
two-mode Gaussian state is not entangled  if $k_x k_p \geq 0$. However, it may
or may not be entangled when $k_x k_p < 0$. A Gaussian state is not entangled
(separable) if, and only if \cite{Sim00}
\begin{equation}
I_1 I_2 + (1 - |I_3|)^2 - I_4 \geq I_1 + I_2,
\label{simon}
\end{equation}
where $I_1 = \mbox{det}(A)$,
$I_2=\mbox{det}(B)$, $I_3=\mbox{det}(C_{\gamma})$ and
$I_4=\mbox{tr}(AJC_\gamma JBJC_\gamma^{T}J)$ are the four invariants of the
$Sp(2,R)\otimes Sp(2,R)$ group \cite{Sim00}.
Here $J$ is an anti-diagonal $2\times 2$ matrix given by
$J=\mbox{adiag}(1,-1)$, $\mbox{det}(M)$ stands for the determinant
of the matrix $M$, $\mbox{tr}(M)$ is the trace of $M$, and $M^{T}$ is the
transpose of $M$.

For a symmetric state ($I_1=I_2$) the entanglement of
formation \cite{Gie03} can be written in terms of the symplectic invariants as
\cite{Rig04}
\begin{equation}
E_f(\rho) = f\left( \sqrt{I_1 + |I_3| - \sqrt{I_4 + 2\,I_1\,|I_3|}}\right),
\label{eof}
\end{equation}
where $f(x)=c_+(x)\log_2(c_+(x)) - c_-(x)\log_2(c_-(x))$ and
$c_{\pm}(x)=(x^{-1/2}\pm x^{1/2})^2/4$. There is no analytical expression of
$E_f$ for non-symmetric Gaussian states.
However, their
entanglement can be quantified
by the negativity, which can be calculated if the above four symplectic
invariants are known \cite{Ade04a,Ade04b}.

\section{The four local invariants}

It is important to note, since this is a crucial ingredient
of our experimental proposal, that the four quantities $I_1$, $I_2$, $I_3$, and
$I_4$ are invariants by local symplectic transformations in the quadratures, or
equivalently, they are invariant by local unitary operations in the density
matrix $\rho$ \cite{Sim00}. The knowledge of these four invariants allows us
to completely characterize the entanglement properties of a two-mode
 Gaussian state:
we can discover if it is entangled (Eq.~(\ref{simon})) as well as how
much it is entangled (Eq.~(\ref{eof})).
Therefore, the measurement of these
symplectic invariants is the base on which our proposal is built and our
goal now is to present an experimental set-up in which they can be
easily determined.

For this purpose it is more appropriate to work with the covariance matrix $V$ 
\cite{footnote4}
of the creation, $a_j^{\dagger}=(X_j-\mathrm{i}P_j)/\sqrt{2}$,
and annihilation,
$a_j=(X_j+\mathrm{i}P_j)/\sqrt{2}$,
operators for the two modes ($j=1,2$) \cite{Oli05}.
Defining $\mathbf{v}=$ $(v_1, v_2, v_3, v_4)^T$ $=$
$(a_1, a_1^\dagger, a_2, a_2^\dagger)^T$, the matrix elements of $V$ are
$V_{ij}=(-1)^{i+j}\langle v_i v_j^\dagger + v_j^\dagger v_i\rangle_{\rho}/2$
\cite{footnote2}.
As we did for $\gamma$ we can represent $V$ in terms of four block
matrices of dimension two:
\begin{equation}
V =
\left(
\begin{array}{cc}
V_1 & C_{V} \\
C_{V}^{\dagger} & V_2
\end{array}
\right).
\end{equation}
In this new representation the four symplectic invariants read:
$J_1 = \mbox{det}(V_1)$, $J_2=\mbox{det}(V_2)$,
$J_3=\mbox{det}(C_V)$ and  $J_4=\mbox{tr}(V_1ZC_VZV_2ZC_V^{\dagger}Z)$, where
$Z=\mbox{diag}(1,-1)$. Recalling
the definitions of $\gamma_{ij}$,
$V_{ij}$, and the relation between $X_j, P_j$ and $a_j, a_j^\dagger$
it is straightforward to see that
\begin{equation}
I_1=4J_1,\, I_2=4J_2,\, I_3=4J_3,\, I_4 = 16 J_4.
\label{iandj}
\end{equation}
Like $\gamma$, under local symplectic transformations $V$ can be brought
to the following standard form \cite{footnote-V}:
\begin{equation}
\tilde{V} =
\left(
\begin{array}{cc}
\tilde{V}_1 & \tilde{C}_{V} \\
\tilde{C}_{V}^{\dagger} & \tilde{V}_2
\end{array}
\right),
\label{Vtilde}
\end{equation}
with
\begin{eqnarray*}
\tilde{V}_1 =
\left(
\begin{array}{cc}
\tilde{n}_1 & 0 \\
0 & \tilde{n}_1
\end{array}
\right), 
\tilde{V}_2 =
\left(
\begin{array}{cc}
\tilde{n}_2 & 0 \\
0 & \tilde{n}_2
\end{array}
\right), 
\tilde{C}_V =
\left(
\begin{array}{cc}
\tilde{m}_s & \tilde{m}_c \\
\tilde{m}_c & \tilde{m}_s
\end{array}
\right),
\end{eqnarray*}
where $\tilde{m}_s$ and $\tilde{m}_c$ are real parameters. From now on, 
whenever we deal with the standard form of $V$, we will write its elements
as well as any related quantity thereof with a tilde ($\tilde{\,}$).

\section{Experimental proposals}

We only need, then, to experimentally measure $J_1$, $J_2$, $J_3$, and
$J_4$ to completely quantify
the entanglement of a two-mode Gaussian state. 
We first show a scheme in which we
can determine the first three invariants. With these three invariants, as 
will be shown in what follows, we can obtain bounds for the 
entanglement of a two-mode Gaussian state. We then introduce two local unitary
operations to the previous scheme, allowing us to determine $J_4$, the 
remaining invariant.
   
\subsection{First scheme}

\begin{figure}[!ht]
\centering
\includegraphics[width=2.75in]{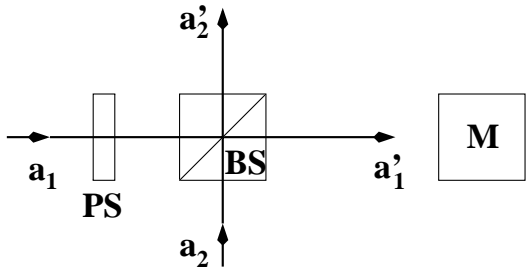}
\caption{\label{fig1} Experimental set-up to measure 
 three symplectic invariants ($J_1$, $J_2$, and $J_3$) of a two-mode 
Gaussian state.
Input modes $a_1$ and $a_2$ pass through a beam splitter (BS) with 
transmittance $\cos \theta$.  Before reaching the BS mode $a_1$ passes 
through a phase-shifter (PS) acquiring a phase $\varphi$. 
After the BS measurements (M) are made on the output mode $a'_1$ only. 
Two types of measurements are required 
to completely characterize the  invariants of the input two-mode Gaussian
state: the photon number and the purity (Wigner function
at the origin of the phase space) of mode $a'_1$.}
\end{figure}

The experimental set-up necessary to measure the first three invariants is 
very simple and is depicted in Fig.~\ref{fig1}. It can be thought of
as a simplification  of the scheme presented in Ref. \cite{Aur05} to
reconstruct the two-mode
covariance matrix of a Gaussian state. Here, however, there is no state
reconstruction and we only use linear
optical devices to measure the purity and the photon number of
the output mode $a'_1$. 
Loosely speaking the linear optics apparatus
(adjustable phase shifter and a beam splitter)
can be seen as the agent responsible for transferring the entanglement
properties between the two modes $a_1$ and $a_2$ to the mode $a'_1$
\cite{Oli05}. On the
other hand, the measuring apparatus (photon counting and/or homodyne detection)
are responsible for the local projective measurements which determine these
properties.
The phase shifter
and the beam splitter action on the modes $a_1$ and $a_2$ is
modeled by the following non-local bilinear Bogoliubov transformation
\cite{Oli05} $\mathbf{v'}=\mathcal{U}\mathbf{v}$, where  $\mathbf{v'}$ $=$
$(a'_1, (a'_1)^{\dagger}, a'_2, (a'_2)^{\dagger} )^T$ and
\begin{equation}
\mathcal{U} =
\left(
\begin{array}{cc}
\mathcal{R} & \mathcal{S} \\
-\mathcal{S}^{*} & \mathcal{R}^*
\end{array}
\right).
\end{equation}
The $2\times 2$ block matrices
$\mathcal{R}=\mbox{diag}(\mathrm{e}^{\mathrm{i}\varphi}\cos \theta,
\mathrm{e}^{-\mathrm{i}\varphi}\cos \theta )$ and
$\mathcal{S} = \mbox{diag}(\sin \theta, \sin \theta )$ are such that
$\cos \theta$
is the transmittance $T$ of the beam splitter and $\varphi$ is the 
phase shift in
mode $a_1$. The two-mode output covariance matrix is $V'=\mathcal{U}^\dagger V
\mathcal{U}$. We will only need, however, the mode $a'_1$ local covariance
matrix, which reads \cite{Oli05}
\begin{equation}
V'_1 = \mathcal{R}^* V_1\mathcal{R} + \mathcal{S}V_2\mathcal{S}^* -
\mathcal{S}C^{\dagger}_V\mathcal{R} -  \mathcal{R}^* C_V\mathcal{S}^* .
\label{v1linha}
\end{equation}

Defining $J'_1(\theta,\varphi) = \mbox{det}(V'_1)$,
where we explicitly write the dependence of $\mbox{det}(V'_1)$ on the
parameters $\theta$ and $\varphi$, and using Eq.~(\ref{v1linha}) we easily see
that
\begin{equation}
J_1 = J'_1(0,0), \hspace{.5cm} J_2 = J'_1\left(\frac{\pi}{2},0\right).
\label{j1andj2}
\end{equation}
As expected $J_1$ and $J_2$ are obtained when the beam splitter has,
respectively, transmittance one ($\theta = 0$) and
reflectivity one ($\theta = \pi/2$). The determination of $J_3$ is not as
trivial, requiring that a set of $J'_1$
measurements be made on the output for
distinct beam-splitter transmittances and phase-shift arrangements, as well as
the measurement of the mode $a'_1$ average photon number
$N'_1(\theta, \varphi) = \langle (a')_1^{\dagger}a'_1 + 1/2 \rangle$.
A straightforward but tedious calculation gives
\begin{equation}
J_3 = \frac{1}{4}\left(\mathcal{J} + \mathcal{N}\right),
\label{j3}
\end{equation}
where
%
%
\begin{eqnarray}
\mathcal{J} &=& J'_1\left(\frac{\pi}{4},0\right) +
J'_1\left(\frac{\pi}{4},\pi\right) +
J'_1\left(\frac{\pi}{4},\frac{\pi}{2}\right) +
J'_1\left(\frac{\pi}{4},-\frac{\pi}{2}\right) \nonumber \\
&&-J'_1\left(0,0\right) - J'_1\left(\frac{\pi}{2},0\right),
\label{j3a} \\
\mathcal{N} &=& [N'_1(0,0)]^2 +
\left[N'_1\left(\frac{\pi}{2},0\right)\right]^2 +
2 \left[N'_1\left(\frac{\pi}{4},0\right)\right]^2
+ 2 \left[N'_1\left(\frac{\pi}{4},\frac{\pi}{2}\right)\right]^2
\nonumber \\
&&- 2 \left[N'_1\left(0,0\right) + N'_1\left(\frac{\pi}{2},0\right) \right]
 \left[N'_1\left(\frac{\pi}{4},0\right) +
N'_1\left(\frac{\pi}{4},\frac{\pi}{2}\right) \right].
\label{j3b}
\end{eqnarray}
%
Eqs. (\ref{j1andj2})-(\ref{j3b}) constitute one of our central results 
and tell that $J_1$, $J_2$ and $J_3$ of the bipartite input state can be 
completely determined by measurements on only one of the beam-splitter 
output ports. This
result contrasts to previous proposals \cite{Ade04a,Ade04b,Fiu04}
where measurements on both modes are always required to obtain
the symplectic invariants, in particular $J_3$. 

Although we do not have yet the fourth invariant, we can obtain a lower bound
for the entanglement of formation as given by Eq.~(\ref{eof}). Noting that 
$J_4$ (or equivalently $I_4$) is a positive quantity and that the function 
$f(x)$ (Eq.~(\ref{eof})) is a decreasing function of $x$ \cite{Gie03} 
we readily obtain the lower bound by setting the fourth invariant to zero. A 
similar approach allows us to derive a bound for the 
negativity \cite{Ade04a,Ade04b}.   

\textit{Remark:}  If, and only if, $V$ is given as or can be brought locally
to either one of the following two distinct forms the previous three
invariants are enough to completely quantify the entanglement of a two-mode
Gaussian state. Indeed, when $\mbox{det}(C_V)\geq 0$ one of the forms reads
$V_1=\mbox{diag}(n_1,n_1)$, $V_2=\mbox{diag}(n_2,n_2)$, and
$C_{V}=\mbox{diag}(m_s, m_s^{*})$. On the other hand, if 
$\mbox{det}(C_V)\leq 0$ the matrix $C_V$ is anti-diagonal, 
$C_{V}=\mbox{adiag}(m_c, m_c^{*})$. Note that we cannot
go from one form to the other via local unitary operations since we have
different signs for $\mbox{det}(C_V)$. A simple
calculation using these two particular covariance matrices gives
\begin{equation}
J_4= 2 |J_3| \sqrt{J_1 J_2}.
\label{j4}
\end{equation}
Hence, as anticipated above, for these two cases $J_1$, $J_2$, and $J_3$ 
are all that is needed to completely characterize the entanglement of a
two-mode Gaussian state. It is worth mentioning, however, that Eq.~(\ref{j4}) 
is only valid for the two special forms of $V$ described above. In general, 
Eq.~(\ref{j4}) is no longer valid. Also, we have not used it in any
calculations that led to the construction of this and the next scheme.

\subsection{Second scheme}

\begin{figure}[!ht]
\centering
\includegraphics[width=2.75in]{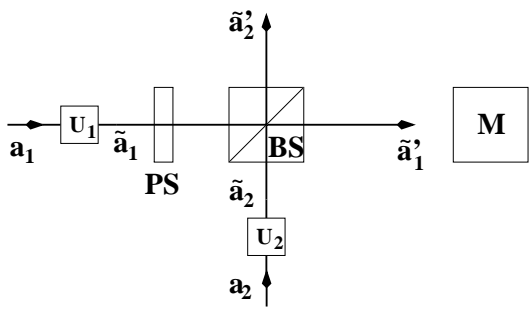}
\caption{\label{fig2} Modified experimental set-up to measure 
 the remaining invariant $J_4$ of a two-mode 
Gaussian state. After the local unitary transformations $U_1$ and $U_2$,
affecting modes $1$ and $2$, respectively, the scheme is identical to the
one given in Fig.~\ref{fig1}.}
\end{figure}

In order to get $J_4$ we modify the previous scheme introducing two local 
unitary operations, as depicted in Fig.~\ref{fig2}.
The unitary transformations $U_1$ and $U_2$ are chosen such that the 
covariance matrix describing the output modes $\tilde{a}_1$ and $\tilde{a}_2$ 
is in its standard form $\tilde{V}$ (See Eq.~(\ref{Vtilde})). These
unitary transformations preserve the Gaussian character of the input state 
and they are equivalent to local symplectic transformations $S_1$ and $S_2$
on $V$. We should
bear in mind that the correct transformation is dependent on the 
one-mode covariance matrices $V_1$ and $V_2$, which fortunately can be 
determined locally by standard homodyne detection. Moreover, the important 
point here is that $S_1$ and $S_2$ (or $U_1$ and $U_2$) always exist and 
that they are local symplectic transformations. 
In the Appendix we show how to express these 
transformations as a function of the input
local covariance matrices $V_1$ and $V_2$.

In the standard form $\tilde{V}$ a direct calculation shows that 
the fourth invariant is given as
\begin{equation}
J_4 = 2 \tilde{n}_1 \tilde{n}_2 \left( |\tilde{m}_s|^2 +
|\tilde{m}_c|^2 \right). 
\end{equation}
Our task then reduces to the determination of $\tilde{n}_1$, $\tilde{n}_2$, 
$|\tilde{m}_s|^2$, and $|\tilde{m}_c|^2$ measuring only the output mode 
$\tilde{a}'_1$, the one obtained after $\tilde{a_1}$ and $\tilde{a_2}$ enter
the beam splitter.  

We first note that $\tilde{n}_1$ and $\tilde{n}_2$ 
are trivially related to invariants already measured in the previous scheme:   
$\tilde{n}_1 = \sqrt{J_1}$ and $\tilde{n}_2 = \sqrt{J_2}$. Second, 
$J_3 = |\tilde{m}_s|^2 - |\tilde{m}_c|^2$ is also known, implying that we just
need to measure either $|\tilde{m}_s|^2$ or $|\tilde{m}_c|^2$ to obtain $J_4$.
Analyzing $\tilde{V}_1$ we see that,
\begin{eqnarray}
|m_c|^2 &=& \left[ \tilde{N}'_1\left(\frac{\pi}{4},0\right) \right]^2 -
\tilde{J}'_1\left(\frac{\pi}{4},0\right),\label{mc2} \\
\mbox{Re}\left(\tilde{m}_s\right) &=& \frac{\tilde{n}_1+\tilde{n}_2}{2} - 
\tilde{N}'_1\left(\frac{\pi}{4},0\right),\\
\mbox{Im}\left(\tilde{m}_s\right) &=& \frac{\tilde{n}_1+\tilde{n}_2}{2} - 
\tilde{N}'_1\left(\frac{\pi}{4},\frac{\pi}{2}\right), \label{imms}
\end{eqnarray} 
where $\tilde{J}'_1(\theta,\varphi) = \mbox{det}(\tilde{V}'_1)$ and 
$\tilde{N}'_1(\theta, \varphi) = 
\langle (\tilde{a}')_1^{\dagger}\tilde{a}'_1 + 1/2 \rangle$ is
mode $\tilde{a}'_1$ average photon number, all quantities determined 
for a given set of parameters $\theta$ and $\varphi$. Note that although 
in the standard form (\ref{Vtilde}) the matrix elements $m_s$ and $m_c$ are 
real, we have assumed them to be complex, which simplifies the unitary
operations $U_1$ and $U_2$ necessary to transform $V$ to $\tilde{V}$. See the
Appendix for more details.

Looking at Eqs. (\ref{mc2})-(\ref{imms}) we see that, similar to the first 
scheme, only two types of measurements are needed to determine 
$J_4$, namely  $\tilde{J}'_1(\theta,\varphi)$ and  
$\tilde{N}'_1(\theta, \varphi)$. 

Additionally, the scheme just presented can be employed without relying on the
first one. Indeed, assuming that we are always dealing with an experimental 
set-up as depicted in Fig.~\ref{fig2}, the first two invariants are
\begin{eqnarray}
J_1 = \tilde{n}_1^2  &=&\left[\tilde{N}'_1(0,0)\right]^2, \\
J_2 = \tilde{n}_2^2  &=& 
\left[\tilde{N}'_1\left(\frac{\pi}{2},0\right)\right]^2.
\end{eqnarray} 
With the aid of Eqs.~(\ref{mc2})-(\ref{imms}), the last two invariants, $J_3$ 
and $J_4$, are readily obtained.

In summary, given a general two-mode Gaussian state the first scheme allows us 
to determine the first three invariants ($J_1, J_2$, and $J_3$) while the 
second scheme gives us the fourth as well as the other three invariants. 
With these four invariants the entanglement content of a two-mode Gaussian 
state is fully determined.

\section{Experimental feasibility}

Our last task is to explain how $J'_1$ and $N'_1$ (or equivalently 
$\tilde{J}'_1$ and $\tilde{N}'_1$ ) are obtained from
experimentally measurable quantities. Firstly, $N'_1$ is
simply the output mode $a'_1$ photon number (or intensity) and can be easily
determined by a photodetection process.
The determinant $J'_1$, on the other hand, is connected to the purity of
mode $a'_1$ by the following expression  \cite{Ade04a,Ade04b}:
$Tr\{(\rho'_1)^2\} = 1/(2\sqrt{J'_1})$.
Moreover, one can prove that \cite{Bra05}  $W(0)=1/(2\pi \sqrt{J'_1})$ where
$W(0)$ is the Wigner function of mode $a'_1$ at the origin of the phase space.
Therefore, any technique developed to measure
the purity and/or $W(0)$ can
be employed to determine $J'_1$. Two interesting proposals were presented in
Refs. \cite{Ban96,Fiu04}. The first one \cite{Ban96},
implemented in Ref. \cite{Ban99b}, shows that
the photon counting statistics allows one to obtain $W(0)$
without any sophisticated data processing. The second
one \cite{Fiu04}, implemented in Ref. \cite{Wen04}, employs only a tunable
beam splitter and a single-photon
detector to obtain the purity of mode $a'_1$. Remark that both schemes do not
require homodyne detection
and permit a direct access to $N'_1$
as well.  However, homodyne detection \cite{Grangier06} can also be employed
for each set of
the parameters $\theta$ and $\varphi$ to reconstruct the output mode $a'_1$
covariance matrix, and thus leading to the immediate calculation of $J'_1$
and $N'_1$.
Nevertheless, 
a complete single-mode reconstruction is more than we need to
fully characterize the entanglement of a two-mode Gaussian state and also 
more experimentally demanding than the previous
two techniques.

In fact what determines the detection scheme to be employed is the specific
scheme detection efficiency in the light frequency range and intensity in use.
Obviously a photon counting scheme for the measurement of the  Wigner function
at the origin of the phase space is more appropriate for low intensity light
fields. On the other hand, homodyne reconstruction is more appropriate 
for continuous bright
light fields, unless an alternative procedure allowing access to the photon
statistics is implemented, such as for pulsed fields detected by on/off
avalanche photodetectors \cite{Zam05}.
Photon counting
(implemented via avalanche devices with saturated gain where each photon
produces a detectable signal (current) at the output) can also be simulated by
dual-homodyne schemes \cite{Nem02}. Such schemes may be useful in infrared and
optical frequencies since the present day avalanche photodiodes possess very
low efficiencies at such communication frequencies \cite{Nem02}.

Whichever the scheme employed imperfect detection and signal losses blur the
measurement outcomes avoiding an exact determination of the invariants as
delineated above. However, in any situation here considered, the losses can be
modeled by a beam splitter of transmittance $\eta$ followed by an ideal
detector, being $\eta$ attributed to the overall efficiency of the detection
scheme \cite{Wen04}. Thus, the actual measured quantities can be corrected by
an appropriate rescaling. Firstly, for homodyne detection the procedure to
measure the detection efficiency $\eta_{hom}$ is well established from
squeezing experiments \cite{Grangier87,Wen04}. Since
$J'_1=I'_1/4=\left(V'_{1,min}V'_{1,max}\right)/4$ and
$N'_1=\left(V'_{1,min}+V'_{1,max}\right)/2$, where $V'_{1,min}$ and
$V'_{1,max}$ are the squeezed and anti-squeezed quadratures variances,
respectively, one can show that in terms of the actual measured variances
${\cal V}'_{1,min}$ and ${\cal V}'_{1,max}$ they are corrected
to $V'_{1,min}=\left({\cal V}'_{1,min}-1+\eta_{hom}\right)/\eta_{hom}$ and
$V'_{1,max}=\left({\cal V}'_{1,max}-1+\eta_{hom}\right)/\eta_{hom}$
\cite{Wen04}.
Secondly, the determination of $J'_1$ and $N'_1$ via photocounting techniques
relies on single photon detection of output mode 1 (given by $a'_1$) after the
two input modes are recombined in a beam splitter of transmittance
$T=\cos\theta$.
The detector efficiency $\eta$ is modeled by a beam splitter of transmittance
$\eta$ followed by an ideal detector, which in Refs. \cite{Fiu04,Wen04}
responds with only two measurement outcomes, while in
Refs. \cite{Ban96,Ban99b} is a number resolving photodetector. In either case
the effects of non-ideal detection can be simply taken into account by
substituting $T\rightarrow \eta T$, whose overall effect is to rescale the
proper transmittance necessary to obtain all the invariants

\section{Conclusion}

We have shown that continuous variable entanglement can be
directly detected and quantified with few simple measurements. In particular,
we have proposed an experimental scheme in which only two types of
measurements, i.e. purity (Wigner function at the origin of the phase space)
and photon number of a single-mode, are required to 
characterize the
entanglement of a two-mode Gaussian state.
Our proposal is valid for either pure or mixed states as
well as for either symmetric or non-symmetric Gaussian states,
giving always the exact value of the state's amount of
entanglement.
Furthermore, 
our scheme can also be seen as a simple procedure to obtain all the
four independent symplectic invariants of a covariance matrix. This allows
one to easily test for the existence of entanglement even in non-Gaussian
states via the Simon's sufficient condition of inseparability.
Finally we remark that the same scheme could be
appropriately adapted for analyzing continuous variable entanglement in other
systems than light fields, such as for the motional degrees of freedom of
trapped ions \cite{Win96}, or even for Bose-Einstein condensates \cite{bruno}.  This is the case whenever similar processes to the
beam splitter and the phase shifter operations can be applied and whenever
proper local projective measurements or local state reconstruction can be
implemented. \\

\textit{Note:} After completion of this manuscript we devised a fully 
local procedure by which we can measure the four invariants of a general
 two-mode Gaussian state \cite{Har07}. Contrary to the schemes here presented, 
in Ref. \cite{Har07} there is no need for a non-local unitary operation 
(as given here by the beam splitter). The trade-off is the inclusion of a 
classical communication channel and one-mode parity measurements.

\section*{Acknowledgments}
We acknowledge support from
Fun\-da\-\c{c}\~ao de Amparo \`a Pesquisa do Estado de S\~ao Paulo (FA\-PESP), 
Conselho Nacional de Desenvolvimento Cient\'{\i}fico e Tecnol\'ogico
(CNPq), and Coordena\c{c}\~ao de Aperfei\c{c}oamento de Pessoal de 
N\'{\i}vel Superior (CAPES).

\appendix
\section{Obtaining the standard form $\tilde{V}$}

The covariance matrix $V$ is generally written as:
\begin{eqnarray}
V =
\left(
\begin{array}{cc}
V_1 & C_{V} \\
C_{V}^{\dagger} & V_2
\end{array}
\right) &=&
\left(
\begin{array}{cccc}
n_1 & m_1 & m_s & m_c \\
m_1^* & n_1 & m_c^* & m_s^*\\
m_s^* & m_c & n_2 & m_2 \\
m_c^* & m_s & m_2^* & n_2
\end{array}
\right),
\end{eqnarray}
where $V_j$, $j=1,2$, and $C_V$ are block matrices of dimension
two. Unitary local operations preserving the Gaussian character of
a state described by $V$ are mapped to the following local
symplectic transformation \cite{Sim94}:
\begin{equation}
S = \left(
\begin{array}{cc}
S_1 & 0 \\
0 & S_2
\end{array}
\right),
\end{equation}
where $S_j$, $j=1,2$, is given as
\begin{equation}
S_j = \left(
\begin{array}{cc}
\mathrm{e}^{-i\alpha_j}\cosh\theta_j&
\mathrm{e}^{i\beta_j}\sinh\theta_j \\
\mathrm{e}^{-i\beta_j}\sinh\theta_j&
\mathrm{e}^{i\alpha_j}\cosh\theta_j
\end{array}
\right),
\end{equation}
with $\alpha_j$, $\beta_j$, and $\theta_j$ real parameters. The
first one is related to rotations of the quadratures of the
quantized electromagnetic field and the last one to local
squeezing operations. The new covariance matrix $\tilde{V}$ is
connected to $V$ by the following relation \cite{Sim94,Oli04},
\begin{equation}
\tilde{V} = S V S^{\dagger},
\end{equation}
which implies that
\begin{eqnarray}
\tilde{V}_j &=& S_j V_j S_j^{\dagger}, \label{newV}\\
\tilde{C}_V &=& S_1 C_VS_2^{\dagger}.
\end{eqnarray}
Explicitly, Eq.~(\ref{newV}) gives for the non-diagonal term of
$\tilde{V}_j$:
\begin{eqnarray}
\tilde{m}_j &=&
\mathrm{e}^{-2\mathrm{i}\alpha_j}m_j\cosh^2\theta_j
+\mathrm{e}^{2\mathrm{i}\beta_j}m_j^*\sinh^2\theta_j \nonumber\\
&&+ \mathrm{e}^{\mathrm{i}(\beta_j-\alpha_j)}n_j\sinh(2\theta_j).
\end{eqnarray}
By setting $m_j=|m_j|\mathrm{e}^{\mathrm{i}\mu_j}$ and solving
for $\tilde{m}_j=0$ we get
\begin{eqnarray}
\alpha_j=\beta_j=\frac{\mu_j+\pi}{2}, \\ 
\tanh(2\theta_j) =\frac{|m_j|}{n_j}.
\end{eqnarray}
Note that this is one of several other solutions. However, for our
purposes, one is enough to prove that $V$ can be locally
transformed to $\tilde{V}$.


\begin{thebibliography}{50}
\bibitem{footnote1} From now on we assume $\langle R_i \rangle_{\rho} = 0$
since the first moments does not affect the entanglement content of a CV-state
and can locally be set to zero by a displacement of the $X$ and $P$
quadratures.
\bibitem{footnote4} We call $V$ the covariance matrix for the creation and 
annihilation operators while $\gamma$ is the covariance matrix of 
the quadratures of the electromagnetic field.
\bibitem{footnote2} Note that since we assumed $\langle R_i \rangle_{\rho} = 0$
we also have  $\langle a_i \rangle_{\rho} = 0$.
\bibitem{footnote-V} Starting with the standard form of $\gamma$ one can
easily see that $V$ can be brought to (\ref{Vtilde}) by expressing it as a 
function of the matrix elements $\gamma_{ij}$.
\end{thebibliography}
\end{document}